\documentclass[12pt]{article}
\usepackage[dvips]{graphicx}
\usepackage{amssymb}
\title{Asymptotic expansions of unstable (stable) manifolds in
 time-discrete systems}
\author{Shin-itiro Goto and Kazuhiro Nozaki\\
{\it Department of Physics,Nagoya University,Nagoya 464-8602,Japan}}
\date{}
\makeindex

\begin{document}
\maketitle

\newcommand{\beq}{\begin{equation}}
\newcommand{\beqa}{\begin{eqnarray}}
\newcommand{\eeq}{\end{equation}}
\newcommand{\eeqa}{\end{eqnarray}}
\newcommand{\non}{\nonumber}
\newcommand{\lb}{\label}
\newcommand{\fr}[1]{(\ref{#1})}
\newcommand{\tx}{\tilde{x}}
\newcommand{\tg}{\tilde{g}}
\newcommand{\hx}{\hat{x}}
\newcommand{\tA}{\tilde A}
\newcommand{\tB}{\tilde B}
\newcommand{\tc}{\tilde c}
\newcommand{\btA}{\mbox{\boldmath {$\tilde A$}}}
\newcommand{\bA}{\mbox{\boldmath {$A$}}}
\newcommand{\bu}{\mbox{\boldmath {$u$}}}
\newcommand{\bN}{\mbox{\boldmath {$N$}}}
\newcommand{\bZ}{\mbox{\boldmath {$Z$}}}
\newcommand{\bR}{\mbox{\boldmath {$R$}}}

\begin{abstract}
By means of an updated renormalization method, we construct
asymptotic expansions for unstable manifolds of hyperbolic
fixed points in the double-well map and the dissipative H\'enon map, both
of which exhibit the strong homoclinic chaos. In terms of the
asymptotic expansion , a simple formulation is presented to give
 the first homoclinic point in the double-well map.
Even a truncated expansion of the unstable manifold is shown to
reproduce the well-known many-leaved (fractal) structure of the strange
 attractor in the H\'enon map.

\end{abstract}

\section{Introduction}
\qquad
The bifurcation of separatrices and homoclinic or heteroclinic structures are
well known to lead to genesis of chaos in conservative dynamical
systems\cite{Poincare}.
While numerical iterations of low-dimensional mappings easily provide these
complicated structures, it is extremely difficult to derive them
analytically.
For nearly integrable systems, the difficulty may have been overcome by
means of the asymptotic expansion beyond all orders supplemented with the
 Borel sum and the stokes phenomenon \cite{Lazutkin}
 \cite{Nakamura} \cite{Gelfreich} \cite{Hirata}.
However, this method can not  be applied to the systems far
from integrable such as a finite-time-discrete dynamical system
with double-well potential (the double-well map) and the strange
attractor in the dissipative H\'enon map,
which will be analyzed in the present paper.\\
\qquad
Recently a novel method based on the perturbative renomalizaion group
theory has been developed as an asymptotic singular perturbation technique
 \cite{Chen1996}.
The renormalization group (RG) method removes secular or divergent terms
 from a perturbation series by renormalizing
integral constants of lower order solutions.
This RG method was reformulated on the basis of a naive renormalization
transformation and the Lie group \cite{Goto}.
The application of the RG method to some non-chaotic discrete systems was
achieved in the framework of the envelope method \cite{Kunihiro}.
In this paper, we apply the reformulated RG method to
chaotic discrete systems.
 Our approach is straightforward and much simpler
 than the original RG method \cite{Chen1996} or the envelope method
 \cite{Kunihiro} and we obtain
asymptotic expansions of unstable and stable manifolds constituting 
homoclinic tangles in the double-well map  and an unstable
manifold included in the closure of the strange attractor of
the dissipative H\'enon map. 
\section{Double well map}
\qquad
Let us  analyze a symplectic map obtained by time-discretization of
canonical equations for dynamical systems with a single freedom:
\beq
x_{n+1}-x_{n}=p_{n+1},\qquad p_{n+1}-p_{n}=\delta^2 f(x_n). \lb{dw-x} 
\eeq
For concreteness, we choose a double well potential
$$ 
f(x_n)=x_n -2 {x_n}^{3}.   
$$ 
Then we get the  second-order difference equation for $x_{n}$
\beq
Lx_n\equiv x_{n+1}-(2+\delta^2)x_n +x_{n-1}
=-2\delta^2 {x_n}^3, \lb{dw}
\eeq
where $\delta^2$ represents the time difference and is not a small
parameter in general. This map is area-preserving and has a hyperbolic
fixed point at $(x,p)=(0,0)$. Although we concentrate on the
system with a double-well potential in this section,
the following analysis will hold for more general systems such as
the standard map.
For $\delta^2 \to 0$ (i.e. in the continuous limit), the system becomes
integrable and the phase space is
occupied by regular orbits separated by a separatrix.
If  $\delta^2\ne 0$, the splitting of
separatrix occurs and unstable and stable manifolds of the hyperbolic
fixed point cross each other at an infinite number of points (homoclinic
points) that accumulate to the hyperbolic fixed point. This complex
homoclinic structure (homoclinic tangles) is the typical chaos generated by
the Birkhoff-Smale's horse shoe mechanism.
When $\delta^2$ is small enough, the splitting of separatrix is
very small and an asymptotic expansion  of unstable
and stable manifolds was obtained by means of the singular perturbation
method of beyond all order \cite{Nakamura}. Here, we are interesting
in the case that $\delta^2\sim {\cal O}(1)$ or larger.
Let us construct a formal series solution $x^{u}_{n}$ near
the hyperbolic fixed point $(x,p)=(0,0)$ along the unstable manifold, that is,
$\mbox{lim}_{n \to -\infty} x^{u}_{n}=0$:
(For brevity, $x^{u}_{n}$ will be noted as $x_{n}$ in the following.)
\beq
x_n=x^{(1)}_n+x^{(2)}_n+x^{(3)}_n+x^{(4)}_n+\cdots. \lb{s-solution}
\eeq
Here $\delta^2$ is not necessarily small and
 $|x^{(1)}_n|>|x^{(2)}_n|>\cdots$.
Then, naive perturbed equations for $x^{(j)}_n$ are written as
\beqa
Lx^{(1)}_n&=&0, \non\qquad\qquad\quad
Lx^{(2)}_n=0,\non\\
Lx^{(3)}_n&=&-2\delta^{2}x^{(1)3}_{n}, \non ~\quad
Lx^{(4)}_n=-6\delta^{2}x^{(1)2}_{n}x^{(2)}_{n}, \non\\
\cdots .\non
\eeqa
 The leading-order solution satisfying
the condition $\mbox{lim}_{n \to -\infty} x_n=0$ is given by
\beq
x^{(1)}_{n}=AK^{n},\lb{dw-leading}
\eeq
where $A$ is an integral constant,
$\delta=2\sinh (k/2)>0$ and $K=\exp{k}>1$.
The series solution \fr{s-solution} is obtained as
\beq
x_{n}=AK^{n}\Bigl(1-{c_2}A^{2}K^{2n}+c_4A^{4}K^{4n}
-c_6A^{6}K^{6n}+\cdots\Bigr),\lb{dw-sol}
\eeq
where
 $c_2$, $c_4$ and $c_6$ are constants defined as
\beqa
c_2&=&\frac{{\delta}^{2}}{D_3},\qquad\qquad\qquad\qquad
c_4=\frac{3{\delta}^{4}}{{D_3}{D_5}},\non\\
c_6&=&\frac{3{\delta}^{6}}{{D_7}{D_3}}
    \bigl(\frac{1}{D_3}+\frac{3}{D_5}\bigr),\quad
D_j=\cosh jk -\cosh k \quad(j\in \bN). \non
\eeqa
Now we introduce a renormalized variable $\tA(n)$ so that
all the higher order terms in the series solution \fr{dw-sol}
are renormalized into the integral constant $A$, that is, 
\beq
x_{n}=\tA(n)K^{n},\lb{dw-solu}
\eeq
where
\beq
\tA(n)=A\Bigl(1-c_2A^{2}K^{2n}+c_4A^{4}K^{4n}
-c_6A^{6}K^{6n}+\cdots \Bigr), \lb{dw-tA}
\eeq
which is called a renormalizaion transformation $A\to \tA$ \cite{Goto}.
In order to derive a difference equation (a renormalizaion equation)
for the renormalized variable $\tA(n)$, we express a difference
 $\tA(n+1)-\tA(n)$ in terms of $\tA$.
 From \fr{dw-tA}, we have
 \beqa
\tA(n+1)-\tA(n)&=&
-c_2(K^2-1)K^{2n} A(n)^3\non \\
&&+c_4(K^4-1)K^{4n}A(n)^{5}+\cdots. \lb{dw-Ad}
\eeqa
By solving \fr{dw-tA} with respect to $A$ iteratively,
$A$ can be expressed in terms of $\tA$ and $n$ as
$$
A=\tA(n)\{1+c_2\tA(n)^{2}K^{2n}+(3c_2^{2}
-c_4)\tA(n)^{4}K^{4n}+\cdots\}.\non
$$
Thus we can replace $A$ in \fr{dw-Ad} by $\tA$ and obtain a
renormalizaion equation:
\beqa
\tA(n+1)-\tA(n)&=&
-c_2(K^2-1)K^{2n}\tA(n)^3\non \\
&&+\{c_4(K^4-1)-3c_2^{2}(K^{2}-1)\}
K^{4n}\tA(n)^{5}\non \\
&&-\{c_6(K^6-1)-5c_2c_4(K^4-1)\non \\
&&+3c_2(K^2-1)
 (4c_2^2-c_4)\} K^{6n} \tA(n)^{7}+\cdots,
 \lb{dw-rn}
\eeqa
which is a non-autonomous difference equation.
In terms of the original variable $x_{n}$ (see \fr{dw-solu}),
the renormalizaion equation \fr{dw-rn} is transformed into
an autonomous one.
\beqa
x_{n+1}&=&Kx_{n}-Kc_2(K^2-1)x_n^3 \non \\
&&+K\{c_4(K^4-1)-3c_2^2(K^2-1)\}x_n^5 \non \\
&&-K\{c_6(K^6-1)-5c_2c_4(K^4-1)\non\\
&&~+3c_2(K^2-1)(4c_2^2-c_4)\}x_n^7 +\cdots \non\\
&\equiv&g_1(x_n),\lb{dw-rg1}
\eeqa
which gives the first branch $g_1$ of a return map on the unstable manifold:
 $x_{n+1}=g_j(x_{n})$, where $g_j$ is a multi-valued function of $x_n$
 and $j(=1,2,\cdots)$ denotes a branch number.
 When $K\gg 1$ (strong chaos) ,
 it is easy to see that
 \beq
 g_1(x)=K[f(x)+{\cal O}(K^{-1})],\lb{s-chaos}
 \eeq
 in general. If $(K-1)\ll 1$ (weak chaos), we have,
 \beq
 g_1(x)=x+(K-1)x(1-x^2)^{1/2}+{\cal O}((K-1)^2). \lb{w-chaos}
 \eeq
for the double well map for
$K=2.1$,
an asymptotic expansion of $g_1$ truncated at $x_n^7$ is depicted
 in Fig .1.
 Although the truncated expression agrees well
 with the exact numerical result for $x_n\lesssim 0.75$, it deviates
 considerably from the exact result near a singular point or a branch point
($x_n\approx 1.1  $) .
In order to avoid this considerable discrepancy near a branch point,
we restrict the domain of $g_1$ so that the return map $g_1$ is
reversible, that is, $x_n\lesssim 0.75$.  The reversible branch thus obtained
is denoted by  $\tg_1$ .
Then, from \fr{dw}, we have the following functional
map: $\tg_{j} \to \tg_{j+1}$
\beq
\tg_{j+1}(x_{n})
=2x_{n}-\tg_{j}^{-1}(x_{n})+\delta^{2}f(x_{n}),\lb{dw-rgj}
\eeq
where the domain of $x_n$ is chosen for each $j$ so that
a new branch $\tg_{j+1}$ is reversible.
Using the functional map \fr{dw-rgj}, we can construct each reversible
branch of the return map $\tg_j$  from the first branch $\tg_1$ or
$g_1$ step by step.
As shown in the figures 2 and 3, the result agrees well with the exact
one even if a truncated expansion of $g_1$ up to $x_n^7$ is used for $\tg_1$.
Once the return map $\tg_j$ on the unstable manifold is constructed,
the unstable manifold $(x^u,p^u)$ is written as
\beq
p^u=x^u-\tg_j^{-1}(x^u), \lb{dw-uns}
\eeq 
because $p^{u}_{n}=x^{u}_{n}-x^{u}_{n-1}=
x^{u}_{n}-\tg^{-1}_{j}(x^{u})$.\\

Using the symmetry of the map \fr{dw-x}:
\beq
x'=x-p,\qquad p'=-p, \lb{sym}
\eeq
the stable manifold $(x^s,p^s)$ is constructed as follows.
The symmetry \fr{sym} gives
\beq
x^u=x^s-p^s,\qquad p^u=-p^s. \lb{sym1}
\eeq
From \fr{sym1} and \fr{dw-uns},
we obtain the stable manifold in terms of $\tg_j$
\beq
p^s=x^s-\tg_j(x^s). \lb{dw-sta}
\eeq
Therefore, the x-coordinate of a homoclinic point $(x^h,p^h)$ is  given by
\beq
x^h=\tg_i\tg_j(x^h).
\eeq
Since the first intersection point $(x^* ,p^* )$
is located at $p^* =0$ \cite{Gelfreich} on the first branch $g_1$ or $\tg_2$
in the present case, $x^*$ is a fixed point of the first branch of the
return map.
\beq
x^*=g_1(x^* )=\tg_2(x^* ). \lb{fix}
\eeq
For $K\gg 1$ and $(K-1)\ll 1$, \fr{s-chaos}, \fr{w-chaos} and \fr{fix}
yield $x^* \approx\pm 1/\sqrt{2}$ and $x^* \approx\pm 1$ respectively.
\section{H\'enon map }
\qquad
Here, we construct an unstable manifold in
the strange attractor of  a
dissipative dynamical system known as the H\'enon map:
\beqa
x_{n+1}&=&1-ax_n^2+y_n, \lb{henon-x} \\
y_{n+1}&=&b x_n,        \lb{henon-y}
\eeqa
where $a=1.4$ and $b=0.3$. This map has the hyperbolic fixed points:
\beqa
x^{*}&=&\frac{b-1\pm \{ {(b-1)}^2+4a\}^{1/2}}{2a}, \non\\
y^{*}&=&b{x}^{*}.                                  \non
\eeqa
Let us consider the unstable manifold of one of the fixed points:
$x^{*}=[b-1+\{ {(b-1)}^2+4a\}^{1/2}]/2a$, $y^{*}=b{x}^{*}$.
This fixed point is known to be included in the closure of the
strange attractor of the H\'enon map.
Setting $\hx_n=x_n-x^*$ and eliminating the y-variable, 
we obtain the second-order difference
equation from \fr{henon-x} and \fr{henon-y}.
\beq
\hx_{n+1}+2ax^{*}\hx_{n}-b\hx_{n-1}=-a \hx_{n}^{2}. \lb{henon}
\eeq
Near the origin of $\hx$, we obtain the following formal series
solution which vanishes as $n \to -\infty$.

$$ 
\hx_n=K^{n}\bigl[ A-c_2 A^2 K^n + c_3 A^3 K^{2n}
    -c_4 A^4 K^{3n}+\cdots \bigl], \non
$$ 
where $A$ is an arbitrary integral constant;
$K=-ax^{*}+\{(ax^{*})^{2}+b\}^{1/2}\approx -1.923739$ is one of the
eigenvalues of the linearized map of  \fr{henon} and
\beqa
c_2&=& \frac{a K^2}{D(K^2)},\qquad \qquad\non
c_3 =\frac{2 a^{2}K^5}{D(K^3)D(K^2)},\non\\
c_4&=& \frac{a^{3}K^8}{D(K^4)}                          
      \{ \frac{1}{D(K^2)^2}+\frac{4K}{D(K^3)D(K^2)} \},\non\\
            D(K^{j})&=&K^{2j}+2ax^{*}K^{j}-b. \non
\eeqa
Following the same procedure as in the previous section,
we renormalize $A$ as
\beqa
\tA(n)&=&A-c_2 A^2 K^n + c_3 A^3 K^{2n}
    -c_4 A^4 K^{3n}+\cdots,  \non \\
\hx_{n}&=&\tA(n)K^{n}, \lb{x-A}
\eeqa
and obtain a similar non-autonomous renormalization equation of $\tA$  as
\fr{dw-rn}. Replacing $\tA(n)$ by $\hx_{n}$ in the
renormalization equation of $\tA$, we have
\beqa
\hx_{n+1}&=&K\hx_{n}-c_2 \hx_n^2 K(K-1) \non \\
&&+\hx_n^3 K(K-1)[c_3(K+1)-2c_2^2] \non \\
&&+\hx_n^4 K(K-1)[-c_4(K^2 +K+1)+3c_3
  c_2 (K+1) \non \\
&&\quad -c_2 (5c_2^2-c_3^3)] \non \\
&&+\hx_n^5 K(K-1)[c_5 (K^{3}+K^{2}+K+1)-4c_2
  c_4(K^2+K+1)\non \\
&&\quad+3c_3 (3c_3^2-c_3)(K+1)
-2c_2 (7c_2^3-6c_2
c_3+c_4)]+\cdots \non\\
&\equiv&g_1(\hx_n),\lb{henon-rg}
\eeqa
which gives the first branch $g_1$ of a return map on the unstable manifold:
 $\hx_{n+1}=g_j(\hx_{n})$, where $g_j$ is a multi-valued function of $\tx_n$
 and $j(=1,2,\cdots)$ denotes a branch number.
 An asymptotic expansion of $g_1$ truncated at $\hx_n^5$ is depicted
 in Fig. 4, where the truncated expression deviates considerably
from the exact numerical result near
$\hx_n\approx 0.7$  and $\hx_n\lesssim -1 $ .
In order to avoid this considerable discrepancy,
we restrict the domain of $g_1(\hx)$ to $-0.7>\hx>0$, where
 the return map $g_1$ is reversible.
The reversible branch in this way obtained
is denoted by  $\tg_1$ .
Then, from \fr{henon}, we have the following functional
map for neighboring reversible branches: $\tg_{j} \to \tg_{j+1}$
\beq
\tg_{j+1}(\hx_{n})=-a\hx^{2}-2a x^{*}\hx_{n}+b\tg_{j}^{-1}(\hx_{n}),\lb{hn-g}
\eeq
where the domain of $\hx_n$ is appropriately chosen for each $j$.
Using the functional map \fr{hn-g}, we construct each reversible
branch of the return map $\tg_j$  from the first branch $\tg_1$
\fr{henon-rg} step by step.
Even when  the asymptotic expansion of $\tg_1$ is truncated at $\hx_n^5$,
the result agrees well with the exact
one as shown in Fig. 5 and Fig. 6.
In Fig. 7, we see how well each branch of the return map constructed from
an initial truncated map $\tg_1$ recovers the well-known
many-leaved (fractal) structure of the strange attractor
near the fixed point \cite{Schuster} .

\section{Conclusion}
\qquad
We construct asymptotic expansions for unstable manifolds of hyperbolic
fixed points of the double-well map and the dissipative H\'enon map, both
of which  exhibit the homoclinic chaos.
Since the dimension of the manifold is one,
the dynamics on the manifolds is described by a return map
with an infinite number of branches.
An asymptotic expansion of the first branch of the
return map is obtained by means of an updated
renormalization method which is an extension of the RG method.
We explicitly give a functional map, through which the other branches are
 obtained consecutively from the first branch.
Thus a global asymptotic form of the unstable manifold is
constructed to give the
homoclinic chaos in the double-well map and the strange attractor in
the dissipative H\'enon map.
Using a truncated expansion of the first branch of the
return map, we calculate an approximate unstable manifold,
 which agrees well with the exact numerical result.
 A fixed point of the first branch of the
return map is found to be the first homoclinic point in the double well map,
and the approximate unstable manifold reproduces the well-known
many-leaved (fractal) structure of the strange attractor in the
H\'enon map.

\section{Acknowledgements}
\qquad
The authors would like to thank Y. Masutomi and other members of R-lab
at Nagoya university for fruitful discussions.


\pagebreak

\begin{figure}[t]
\begin{center}
\includegraphics[width=8cm]{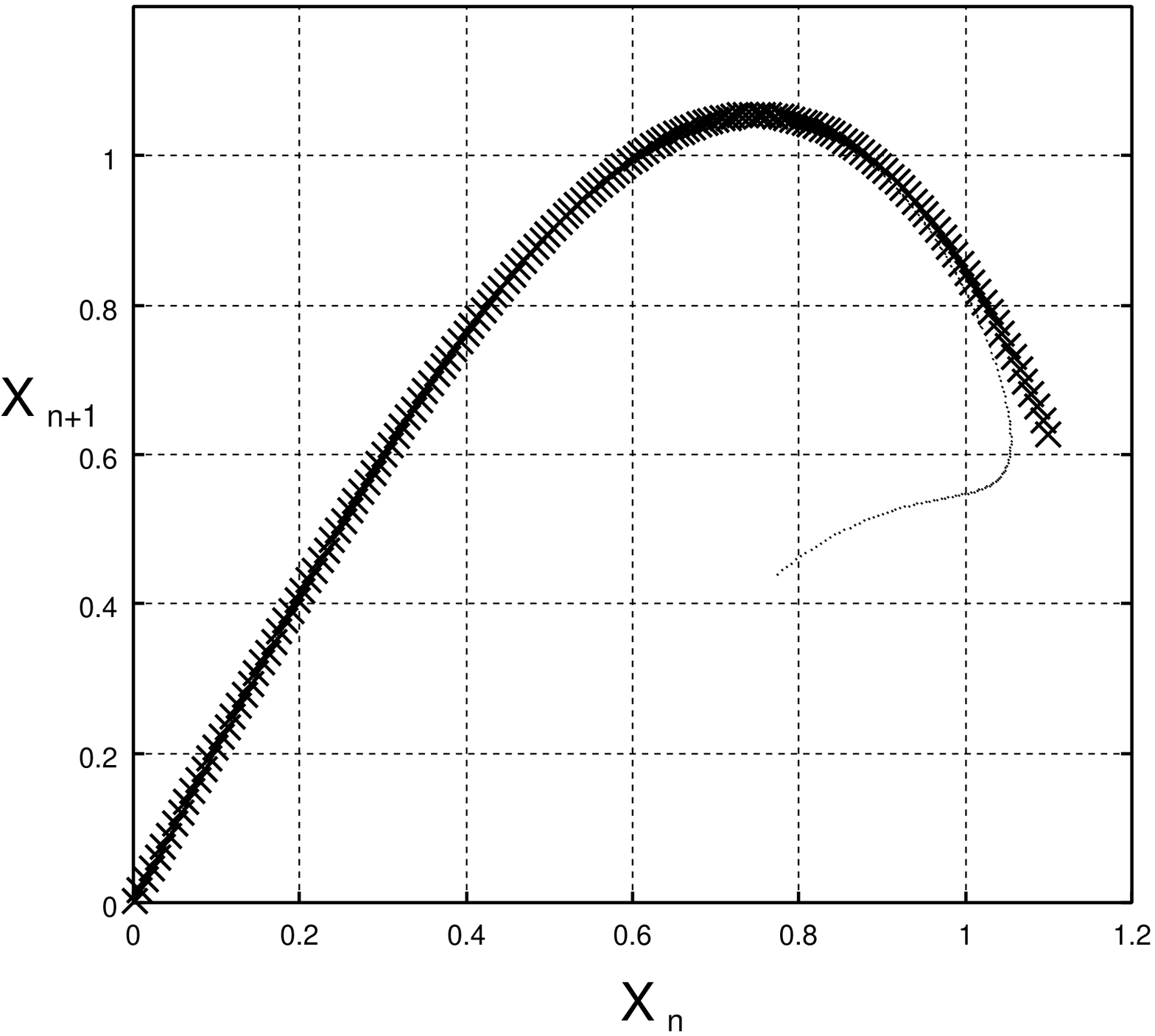}
\caption{ The first branch of the return map on an unstable manifold for the 
double well map. While dots (~$\cdot$~) denote an exact solution 
calculated by means of numerical iterations, pluses (+) denote $\tg_1$ 
obtained by the renormalization method.}
\end{center}
\end{figure}

\begin{figure}[b]
\begin{center}
\includegraphics[width=8cm]{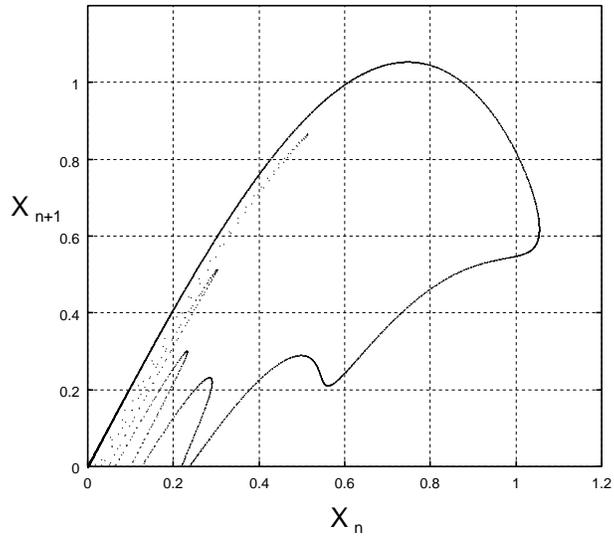}
\caption{The exact return map on an unstable manifold for the double well map.}
\end{center}
\end{figure}

\begin{figure}[t]
\begin{center}
\includegraphics[width=8cm]{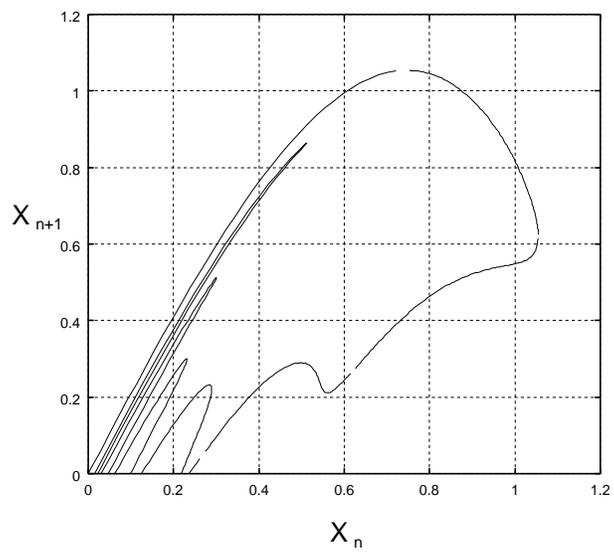}
\caption{The return map plotted by means of matching the set of 
$\tg_j$.} 
\end{center}
\end{figure}

\begin{figure}[b]
\begin{center}
\includegraphics[width=8cm]{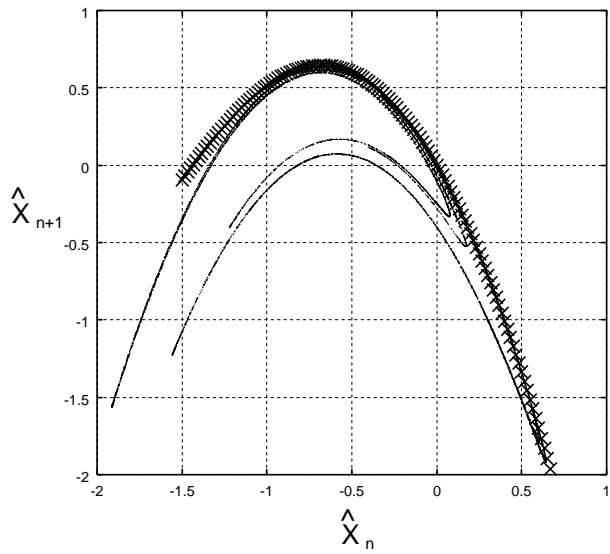}
\caption{ The first branch of the return map on an unstable manifold for the 
double well map. While dots (~$\cdot$~) denote an exact solution 
calculated by means of numerical iterations, pluses (+) denote $\tg_1$ 
obtained by the renormalization method.}
\end{center}
\end{figure}

\begin{figure}[t]
\begin{center}
\includegraphics[width=8cm]{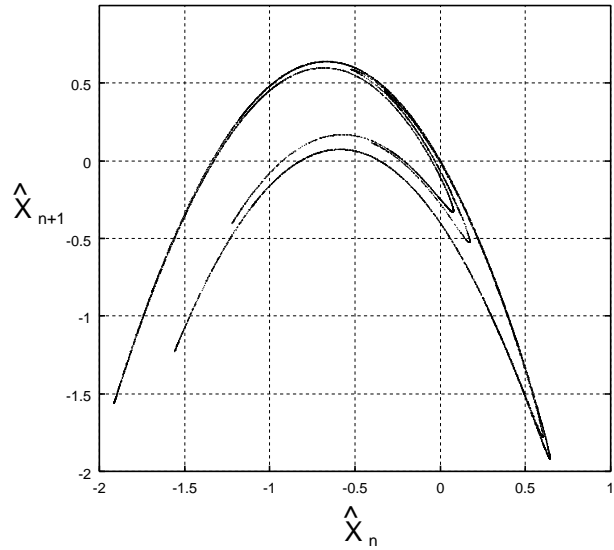}
\caption{The exact return map of the H\'enon map's unstable manifold is 
plotted by numerical iterations. }
\end{center}
\end{figure}

\begin{figure}[b]
\begin{center}
\includegraphics[width=8cm]{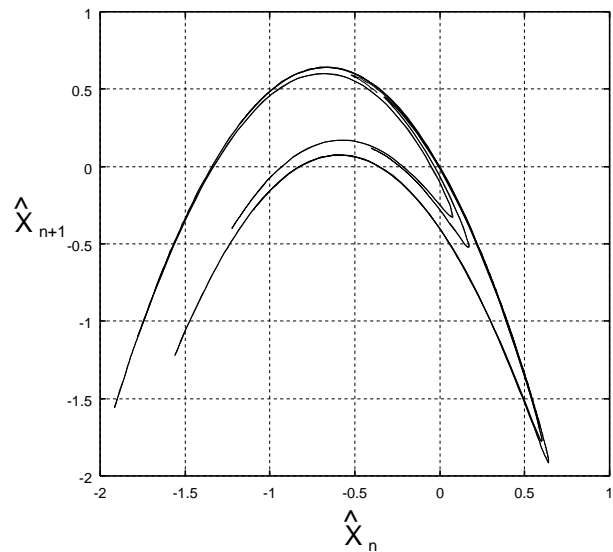}
\caption{ The return map plotted 
by means of the matching the some of the $\tg_j$ .}
\end{center}
\end{figure}

\begin{figure}[t]
\begin{center}
\includegraphics[width=8cm]{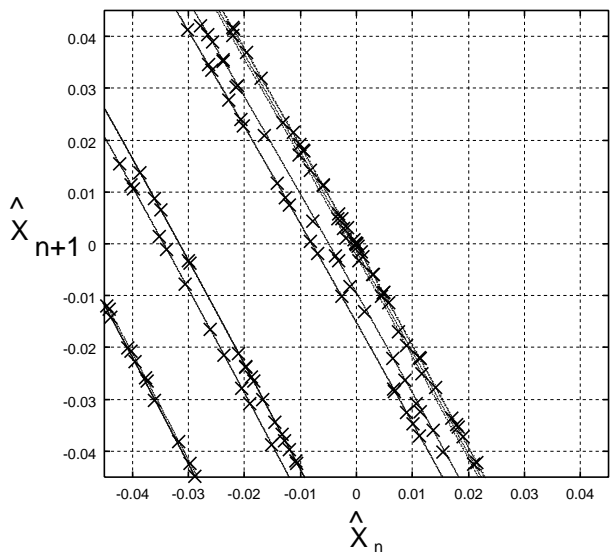}
\caption{The return map of the H\'enon map's unstable manifold 
near the hyperbolic fixed point $(\hx_n ,\hx_{n+1})=(0,0)$ where 
the fractal structure appears.
Pluses (+) express the exact map and
line express the some of the $\tg_j$.}
\end{center}
\end {figure}

\end{document}